\documentclass[12pt, nosubfloats, final]{l4dc2024}

\title[Learning flow functions of spiking systems]{Learning flow functions of spiking systems}
\usepackage{microtype}
\usepackage{times}
\usepackage{subcaption}

\author{%
 \Name{Miguel Aguiar} \Email{aguiar@kth.se}\\
 \addr KTH Royal Institute of Technology, Stockholm, Sweden
 \AND
 \Name{Amritam Das} \Email{am.das@tue.nl}\\
 \addr Eindhoven University of Technology, Eindhoven, Netherlands
 \AND
 \Name{Karl H. Johansson} \Email{kallej@kth.se}\\
 \addr KTH Royal Institute of Technology, Stockholm, Sweden
}

\newcommand{\Par}[1]{\left( {#1} \right)}
\newcommand{\Abs}[1]{\left| {#1} \right|}
\newcommand{\Norm}[1]{\left\| {#1} \right\|}
\newcommand{\Set}[1]{\left\{ {#1} \right\}}
\newcommand{\Bkt}[1]{\left[ {#1} \right]}

\newcommand{\Reals}{{\mathbb R}}

\newcommand{\dd}{\mathrm{d}}        

\newcommand{\Control}{{\mathcal U}} 
\newcommand{\Controls}{{\mathbb U}} 
\newcommand{\State}{{\mathcal X}}   
\newcommand{\Flow}{{\varphi}}       

\newcommand{\Hyp}{{\mathcal H}}     

\newcommand{\E}{{\mathbf E}} 
\newcommand{\Unif}[1]{{\text{Uniform}{\Par{#1}}}}

\newcommand{\Cat}[1]{\underset{{#1}}{\wedge}}

\begin{document}

\maketitle

\begin{abstract}%
We propose a framework for surrogate modelling of spiking systems.
These systems are often described by stiff differential equations
with high-amplitude oscillations and multi-timescale dynamics,
making surrogate models an attractive tool for system design and simulation.
We parameterise the flow function of a spiking system
using a recurrent neural network architecture,
allowing for a direct continuous-time representation of the state trajectories.
The spiking nature of the signals makes for a 
data-heavy and computationally hard training process;
thus, we describe two methods to mitigate these difficulties.
We demonstrate our framework on two
conductance-based models of biological neurons,
showing that we are able to train surrogate
models which accurately replicate the spiking
behaviour.
\end{abstract}


\section{Introduction}

We consider the problem of learning surrogate models of spiking systems from samples
of state trajectories.
Spiking behaviours abound in dynamical system models of biological neurons,
and there is significant interest in reproducing such behaviour in electronic devices.
We propose a data-driven framework based on a recurrent neural network (RNN) architecture
to approximate the flow function of conductance-based state-space models of spiking systems
in continuous time.

Spiking systems~\citep{Sepulchre22} are dynamical systems whose stability
behaviour is highly input-dependent.
Determined by the input excitation, the state typically either remains close 
to an equilibrium  or enters into a limit cycle with large-amplitude oscillations called spikes.
Whether the input excites the system into the oscillatory regime
and how many spikes are emitted depends on both 
the input amplitude and frequency~\citep{SepulchreEtAl18}.
These systems thus possess a mixed continuous--discrete character,
as the spikes can be seen as encoding digital information into a
continuous-time signal.
Models of biological neurons provide prototypical examples of spiking systems,
where the input is a current and the spiking phenomenon is observed in the
membrane potential.

Significant effort in computational neuroscience has been devoted to modelling the spiking
behaviour of neurons.
\citet{HodgkinHuxley52} proposed modelling the relationship between the membrane voltage
and the applied current as a parallel interconnection of nonlinear conductances,
which can be identified using a voltage-clamp experiment, which
\citet{BurghiEtAl21} relate to a general output-feedback system identification scheme.
The parameter identification problem is shown to be tractable
thanks to stability properties of the inverse dynamics of the conductance-based models.

Circuit-theoretic models of biological neurons suggest the possibility of building electronic devices
with spiking behaviour, conceivably combining the best features of analogue and digital electronics 
in a single physical device~\citep{Mead90, DeWeerthEtAl91, Sepulchre22}.
Using these biologically-inspired components in circuit design requires the ability to
efficiently simulate their behaviour, possibly in interconnection with many other circuit elements.
However, the differential equations models of these systems are typically stiff,
which suggests that surrogate models may provide computational advantages over direct integration
of the differential equations.
Furthermore, continuous-time representations are desirable,
due to the possibly input-dependent nature of the periodicity of the spike trains, the
high-frequency nature of the spike signals, and the multiple time scales involved in the dynamics.

Machine learning offers an attractive array of methods for constructing surrogate models
of dynamical systems.
Focusing on continuous-time models,
we can distinguish between two approaches in the literature:
those methods that attempt to learn a model of the system dynamics from data,
and those where the goal is to directly learn a solution operator associated to the system.
In the first class of methods, a neural network is typically used to parameterise the
right-hand side of an ordinary differential equation (ODE),
resulting in a class of models known as neural ODEs.
This requires a way to automatically compute gradients of trajectory values with respect to network
parameters during training, which may be done by differentiating through an ODE solver or using
an adjoint method~\citep{ChenEtAl18}.
In a surrogate modelling context, \citet{YangEtAl22} propose a neural~ODE method 
for learning models of complex circuit elements, and derive a parameterisation 
of the network that guarantees input-to-state stability.
Regarding the application of these models in system identification,
\citet{ForgionePiga21} discuss model structures and fitting criteria,
and \citet{BeintemaEtAl23} propose an architecture and estimation method
shown to compete with state-of-the-art nonlinear identification methods.
The second group of methods, broadly known as operator learning,
attempt instead to directly learn the solution map of a differential equation,
i.e. the mapping from initial conditions, parameters, and external inputs to the solution.
Directly parameterising the solution allows for fast evaluation for new inputs,
and provided that standard deep learning toolchains are used,
gradients with respect to the inputs of the model also become easy to compute.
A great deal of research attention has focused on learning solution operators of
partial differential equations using integral kernel parameterisations
composed with neural networks~\citep{LuEtAl21, LiEtAl21, KissasEtAl22}.
\citet{LinEtAl23} use one such parameterisation in a recursive architecture to
predict trajectories of dynamical systems with external inputs.
In a similar context, \citet{QinEtAl21} suggest a residual network-based architecture
to approximate the one-step-ahead map of a dynamical system.
\citet{BilosEtAl21} suggest a number of architectures inspired by flow functions of
autonomous systems as a substitute for neural ODEs.

Our main contribution is an operator learning framework for constructing continuous-time 
input--output surrogate models of spiking systems from samples of state trajectories.
Starting from the flow function description of the spiking system,
we directly parameterise its state trajectories.
This is particularly suited for systems exhibiting spiking behaviour,
as it allows for a continuous-time description of the state trajectories,
while not requiring sampling the derivatives of the states
(which due to the spikes can widely vary in amplitude).
Indeed, from the flow function point of view
the problem can be formulated as a standard (albeit infinite-dimensional) regression problem.
Furthermore, by imposing the non-restrictive assumption that the input signal is piecewise constant,
we show that there is an exact correspondence between the flow function and a discrete-time
dynamical system.
This suggests that one can approximate the flow function by an RNN,
resulting in an architecture that uses only standard learning components and can be
easily implemented and trained with established deep learning toolchains.
Due to the nature of the spike signals, the trajectories must be densely sampled
in order to correctly capture the timing and height of each spike.
We propose a simple data reduction method based on rejection sampling,
which enables a significant reduction of the required amount of time samples per trajectory
by focusing on the most important regions of the signal.
This is made possible by our continuous-time approach,
which naturally allows for data that is irregularly sampled in time.
Moreover, we show how the complexity of the optimisation problem can be
subdued by considering segments of trajectories, using the properties of the
flow function.
We numerically evaluate our approach through simulations of
conductance-based models of a single neuron and an interconnection of two neurons.
We build upon our recent work~\citep{AguiarEtAl23a} in which we describe a
neural network architecture for learning flow functions of continuous-time control systems.
In the current paper, we focus explicitly on challenges arising 
in learning the dynamics of spiking systems
which require special attention when applying this architecture.

The remainder of the paper is organised as follows.
In section~\ref{sec:prob-form} we introduce conductance-based neuron models
in state-space form as a prototype for spiking behaviours,
define the concept of flow function of a control system,
and formulate the problem of constructing a surrogate model of such a system
as an optimisation problem.
In section~\ref{sec:methodology} we describe the proposed architecture and
two methods for reducing the complexity of the training process.
Finally, in section~\ref{sec:experiments} we report and discuss results
from numerical experiments, and in section~\ref{sec:conclusion} we conclude
and discuss possibilities for future work.

\section{Problem formulation}\label{sec:prob-form}
We begin by introducing a class of state-space models of biological neurons 
and their interconnections which serve as prototypical examples of systems
exhibiting spiking behaviour,
before introducing the definition of the flow function of a control system
and the mathematical formulation of the surrogate modelling problem for these systems.

\subsection{Conductance-based models}
The general conductance-based model of a neuron is given by
the system of differential equations~\citep{BurghiEtAl21}
\begin{equation}\label{eq:cond-model}
\begin{aligned}
    C_m\dot{V}(t) &= u(t) - g_{\text{leak}}(V(t) - V_{\text{leak}}) - \sum_{k = 1}^{N_I} I_k(V(t), m_k, n_k) \\
    \dot{m}(t) &= A_m(V(t))m(t) + b_m(V(t)) \\
    \dot{n}(t) &= A_n(V(t))n(t) + b_n(V(t)),
\end{aligned}
\end{equation}
where $V$ is the membrane potential, $u$ the external (input) current, $C_m > 0$ the membrane capacitance,
$g_{\text{leak}}$ the leak conductance, and $V_{\text{leak}}$ the reversal potential.
The gating variables $m, n \in [0, 1]^{N_I}$ are dimensionless,
and $A_m(V), A_n(V)$ and $b_m(V), b_n(V)$ are diagonal matrices
and vectors depending on $V$, respectively.
The ionic currents are given by
$I_k(V, m_k, n_k) = g_k m_k^{\alpha_k} n_k^{\beta_k} (V - V_k)$,
where $g_k > 0$ and $V_k \in \Reals$ are constants 
and $\alpha_k, \beta_k$ are nonnegative integers.
A detailed description of this class of models and their biological motivation is given 
in~\citet{HodgkinHuxley52} and~\citet{PospischilEtAl08}.
These models are prototypes of spiking systems:
for certain choices of the input current one observes spiking behaviour
in the membrane potential signal $V$~\citep{SepulchreEtAl18}.

One can interconnect several neuron models~\eqref{eq:cond-model} to obtain more complex
spiking behaviours~\citep{GiannariAstolfi22}.
In particular, we can model the interconnection of $n$ neurons through electrical synapses
\begin{equation}\label{eq:cond-model-multi}
    C^i_m\dot{V}_i(t) = u_i(t) - g^i_{\text{leak}}(V_i(t) - V^i_{\text{leak}}) -
        \sum_{k = 1}^{N^i_I} I^i_k(V_i(t), m^i_k, n^i_k)%
        + \sum_{j = 1}^n{ \epsilon_{ij}(V_j(t) - V_i(t)) },
\end{equation}
where $m^i$ and $n^i$ satisfy differential equations with the same structure as those
in~\eqref{eq:cond-model} (depending only on $V_i$)
and $\epsilon_{ij} \geq 0$ is the weight of the electric synapse from neuron $j$ to neuron $i$.

\subsection{Flow functions}

Consider a dynamical system described in state-space form by
\begin{equation}\label{eq:ode}
\begin{aligned}
    \dot{\xi}(t) &= f(\xi(t), u(t)), 
    ~ \xi(0) = x \\
    \eta(t) &= h(\xi(t)),
\end{aligned}
\end{equation}
with state $\xi(t) \in \Reals^{d_x}$, input $u(t) \in \Reals^{d_u}$ and output $\eta(t) \in \Reals^{d_y}$.
Assume that $f$ is such that solutions to~\eqref{eq:ode} exist on $\Reals_{\geq 0}$ 
for $x \in \State$, with $\State \subset \Reals^{d_x}$ being an invariant set,
and $u: \Reals_{\geq 0} \to \Control$ is measurable and essentially bounded.
Define a map ${\Flow: \Reals_{\geq 0} \times \State \times \Controls \to \State}$,
$\Controls := L^{\infty}(\Reals_{\geq 0}, \Control)$,
such that for any such $x$ and $u$ it holds that $\xi(t) = \Flow(t, x, u), ~ t \geq 0$.
The map $\Flow$ is called the flow function of~\eqref{eq:ode}.
The flow function satisfies the identity property:
for any $x \in \State$ and $u \in \Controls$, $\Flow(0, x, u) = x$;
and the semigroup property: for any $x \in \State$, $t, s \geq 0$ and $u, v \in \Controls$,
$\Flow(t + s, x, u\Cat{s}v) = \Flow(t, \Flow(s, x, u), v)$.
Here $u \Cat{s} v$ denotes the concatenation of $u$ and $v$
at time $s \geq 0$, defined as $[{u \Cat{s} v}]{(t)} = u(t)$, $0 \leq t < s$, and
$[{u \Cat{s} v}]{(t)} = v(t - s)$, $t \geq s$.

Henceforth we assume that the considered controls are piecewise constant%
\footnote{
The approach can be extended to more general classes of controls,
but we proceed with this assumption as it simplifies the exposition
and piecewise constant controls are sufficient for our purposes.
See~\citet{AguiarEtAl23b} for a discussion on extensions.%
}
with sampling period $\Delta > 0$.
Thus, let $\Controls^0_\Delta \subset \Controls$, where $\Delta > 0$ and
$u \in \Controls^0_\Delta$ if and only if there exists a sequence
${(\omega_k)_{k = 0}^{\infty} \subset \Control}$ such that
$u(k\Delta + t) = \omega_k$ for all $k \geq 1$ and $t \in [0, \Delta)$.
We let $\Controls_\Delta = \bigcup_{s \in [0, \Delta)} \sigma^s (\Controls^0_{\Delta})$,
where $\sigma$ is the time-shift operator defined by $(\sigma^{s}u)(t) = u(t + s)$
for $s \geq 0$,
and restrict $\Flow$ to $\Reals_{\geq 0} \times \State \times \Controls_\Delta$.

The conductance-based models~\eqref{eq:cond-model} and~\eqref{eq:cond-model-multi}
can be written in the form~\eqref{eq:ode},
with the state $\xi$ given by the membrane potentials $V_i$ together with
the gating variables $m^i_k, n^i_k$ of each neuron and the input signal $u$
given by the collection of input currents $u_i$.
We take the output $\eta$ to be the collection of membrane potentials $V_i$,
as these are the spiking signals we are interested in simulating and,
as in~\eqref{eq:cond-model-multi},
the relevant signals when interconnecting neuron models.

\newpage

\subsection{Problem}
Let us now define the problem considered in this paper.
Let $y(t, x, u) := h(\Flow(t, x, u))$ be the trajectory of the output signal.
We define the problem of obtaining a surrogate model of the system~\eqref{eq:ode}
as that of solving the optimisation problem
\begin{equation}\label{eq:loss}
    \underset{\hat{y}\in\Hyp}{\text{minimise}}~~\ell_T(\hat{y}) := \E_{x, u}~\frac{1}{T}\int_{0}^{T}{
        \Norm{y(t, x, u) - \hat{y}(t, x, u)}_{1}\dd{t}
    },
\end{equation}
where $x, u$ are assumed to be independent and distributed according to probability distributions
$P_x$ and $P_u$, describing the initial conditions and control inputs of interest, respectively.
The distribution $P_u$ has its support in $\Controls_\Delta$,
so that sampling the values of a control $u \sim P_u$ on $[0, T]$ is equivalent
to sampling a finite sequence of control values.
We use the $1$-norm to measure the approximation error due to the sparse nature of the spike signals.
The set
$\Hyp \subset \Set{\hat{y} :
\Reals_{\geq 0} \times \Reals^{d_x} \times \Controls_\Delta \to \Reals^{d_x}}$
is the hypothesis class from which the surrogate model $\hat{y}$ is to be selected.
In this paper, we consider a hypothesis class given by the RNN architecture described in the following 
section.

\section{Methodology}\label{sec:methodology}
In this section, we first motivate the approximation of the flow function by an RNN and
describe the proposed architecture.
We then discuss the data collection process and two issues arising in the training of the RNN architecture,
as well as methods to mitigate them.

\subsection{Architecture}\label{ssec:architecture}
Fix $t \geq 0$, $x \in \State$ and $u \in \Controls^0_\Delta$.
Let $(\omega_k)$ be the sequence of values of $u$,
and define for $\omega \in \Control$ the constant control
$u_{\omega}$ through $u_\omega(t) := \omega$.
With $k_t := \lfloor t/\Delta \rfloor$ (the sample index corresponding to $t$), 
the value of $\Flow(t, x, u)$ can be evaluated recursively as follows:
\begin{equation}\label{eq:flow-recurrence}
    \begin{aligned}
        x_0 &= x \\
        x_{k+1} &= \Flow(\Delta, x_k, u_{\omega_k}), ~ 0 \leq k < k_t \\
        x_{k_t + 1} &= \Flow(t - k_t\Delta, x_{k_t}, u_{\omega_{k_t}}).
    \end{aligned}
\end{equation}
By the semigroup property, we then have $x_{k_t + 1} = \Flow(t, x, u)$.
Letting $\Phi: [0, 1] \times \State \times \Control \to \State$ be defined by
$\Phi(\tau, x, \omega) = \Flow(\tau\Delta, x, u_{\omega})$,
and $\tau_k$ by ${\tau_k = 1}$ for $0 \leq k < k_t$ and 
${\tau_{k_t} = (t - k_t\Delta)/\Delta}$,
we can rewrite~\eqref{eq:flow-recurrence} as
$x_{k+1} = \Phi(\tau_k, x_k, {\omega_k})$, ${0 \leq k \leq k_t}$.

If now $v \in \Controls_\Delta$,
there is some $u \in \Controls^0_{\Delta}$
and $\delta \in [0, \Delta)$ such that $v = \sigma^{\delta}u$,
and by the semigroup property 
$\Flow(t, x, v) = \Flow(t - \delta, \Flow(\delta, x, u_{\omega_0}), \sigma^{\Delta}u)$
for $t \geq \delta$.
Thus, since $\sigma^{\Delta}u, u_{\omega_0} \in \Controls_{\Delta}$,
one can proceed as above also in this case.
This shows that the flow can be exactly computed at any time instant by a discrete-time
finite-dimensional dynamical system (with inputs $(\tau, \omega)$),
suggesting that one can approximate $\Flow$ (and thus $y$) through this representation.

\begin{figure}[!htbp]
    \centering
    \includegraphics[width=0.5\textwidth]{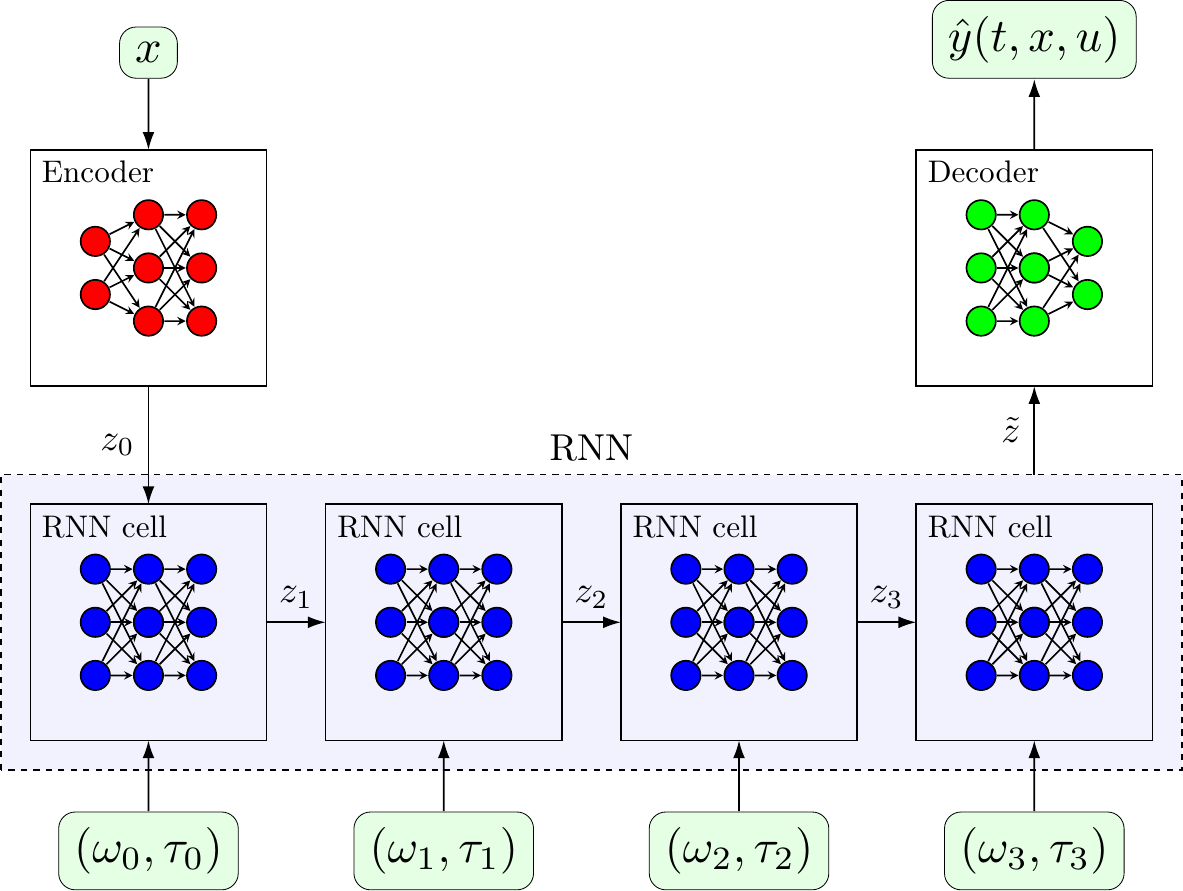}
    \caption{
        Schematic of the proposed architecture for $k_t = 3$.
    }
    \label{fig:arch-diagram}
\end{figure}

The previous derivation motivates the choice of the hypothesis class $\Hyp$ given by a parameterisation of
$\hat{y}$ based on the composition of an RNN with a pair of encoder--decoder networks,
as illustrated in Figure~\ref{fig:arch-diagram}.
This architecture is in fact a universal approximator of the flow function 
corresponding to~\eqref{eq:ode} under mild conditions on $f$~\citep{AguiarEtAl23b}.

\subsection{Data collection}
To generate training data, we integrate $N$ trajectories of 
the differential equation~\eqref{eq:ode}
with initial conditions $x_i$ and $u_i$, $i = 1, \dots, N$,
sampled from $P_x$ and $P_u$, respectively, obtaining samples
\begin{equation}\label{eq:samples}
    \xi_{ik} = \Flow(t_{ik}, x_i, u_i), ~
    k = 1, \dots, K, ~ i = 1, \dots, N,
\end{equation}
where, for each~$i$, the sampling times $t_{ik}$ are uniform on $[0, T]$
and increasing in $k$.
Using these samples, we construct an approximation of $\ell_T$ in~\eqref{eq:loss} as
\begin{equation}\label{eq:emp-loss}
    \hat\ell_T(\hat{y}) := \frac{1}{N}\frac{1}{K}\sum_{i, k} \Norm{h(\xi_{ik}) - \hat{y}(t_{ik}, x_i, u_i)}_1.
\end{equation}

Because $\Flow$ is the flow of a spiking system, the optimisation of~\eqref{eq:emp-loss}
is not without challenges.
The spikes can be very `thin', which can imply that a large number of samples
are required when the sampling times $t_{ik}$ are uniformly distributed,
increasing the computational load during training.
Furthermore, the spikes might be relatively infrequent, and consequently underrepresented in the data,
which can make it harder to learn the spiking behaviour properly.
In the next subsection we describe a simple rejection sampling algorithm that alleviates these issues.

\subsection{Rejection sampling for data reduction}

We propose a method that simultaneously reduces the amount of samples per trajectory 
needed to represent the spike signal and weights the loss function 
in order to emphasise learning the spiking behaviour.
This is done as follows: first, sample data~\eqref{eq:samples} with $t_{ik}$ uniform and dense,
and then use rejection sampling to `prune' the data, i.e.
select which samples to remove so that the remaining $t_{ik}$ 
have a distribution which favours learning the spiking behaviour correctly.

Consider first the case of a single output, i.e. $y$ is scalar-valued.
Roughly speaking, the output signal $y(\cdot, x, u)$
has higher frequency content when its amplitude is higher
(i.e. when a spike is emitted).
One should thus sample more densely when the value of $y(\cdot, x, u)$ is higher.
In other words, we would like that the sampling times $t_{ik}$ be distributed
according to the density function
$p(t, x_i, u_i) \propto \Bkt{y(t, x_i, u_i) - \min_{s \in [0, T]}{y(s, x_i, u_i)}}$
(or more generally $p \propto \alpha(y)$, where $\alpha$ is an increasing function).
If $t_{ik}$ are sampled with this density,
$\hat\ell_T$ in~\eqref{eq:emp-loss} is the empirical estimate of the weighted loss function
\begin{equation*}
    \E_{x, u}~\frac{1}{T}\int_{0}^{T}{
        p(t, x, u)\Norm{y(t, x, u) - \hat{y}(t, x, u)}_1\dd{t}
    },
\end{equation*}
giving higher weight to parts of the trajectories where spiking occurs.

In our case, where $t_{ik}$ are given a~priori and uniformly distributed,
we can use rejection sampling~\citep{Ross13} to discard certain samples
so that the remaining $t_{ik}$ are distributed approximately according to $p$.
This implies the following procedure for each $(i, k)$:
\begin{itemize}
    \item Draw $\Upsilon_{ik} \sim \Unif{[0, 1]}$;
    \item If $\Upsilon_{ik} > \dfrac{p(t_{ik}, x_i, u_i)}{M_i}$,
        where $M_i := \max_{k} p(t_{ik}, x_i, u_i)$,
        remove the sample.
\end{itemize}
After a single pass through the dataset, the undiscarded $t_{ik}$
will be approximately distributed with density
$p(\cdot, x_i, u_i)$.
Furthermore, the probability of accepting a sample from the $i$th trajectory
is approximately equal to $1/M_i$,
giving the approximate fraction of samples which will be retained.


We are thus at once able to reduce the volume of data while retaining a faithful representation
of the spiking signals, and to increase the weight of the spiking regions in the loss function.
Other choices of $p$, e.g. involving the derivative or frequency content of the output signal,
could of course also be used.

If there are several outputs, so $y$ is vector-valued,
one may combine the outputs into a scalar signal that contains the spikes of all outputs,
for instance, $p(t, x, u) \propto \max_{i = 1, \dots, d_y} \alpha_i(y_i(t, x, u))$,
where $\alpha_i$ are monotone functions to ensure that $y_i$, $i = 1, \dots, d_y$,
are normalised to the same range.

\subsection{Windowed loss using the semigroup property}\label{ssec:semigroup}

The complexity of optimising the empirical loss~\eqref{eq:emp-loss}
with an RNN is highly dependent on the length of the input sequences.
This dependence is twofold: the computational effort of the forward and backward passes
through the recurrent network depends linearly on the simulation length,
while simultaneously the loss function becomes less smooth with respect
to the network parameters as the sequence length increases~\citep{RibeiroEtAl20}.
We describe here how this issue can be addressed in the context of our method,
by reducing the length of input sequences while still making use of
all training data.
In a similar way to~\citet{RibeiroEtAl20} and~\citet{BeintemaEtAl23},
we construct a new loss function by considering shorter segments of output trajectories.
This is easy to do in our setting, as we can take advantage of properties
of the flow function.

It follows from the semigroup property that for each $i, k$ and $j \geq k$ we can rewrite
each trajectory sample $\xi_{ij}$ given as in~\eqref{eq:samples} as
${\xi_{ij} = \Flow(t_{ij} - t_{ik}, \xi_{ik}, \sigma^{t_{ik}}u_i)}$.
Note that $\sigma^{t_{ik}}u_i \in \Controls_\Delta$,
so that with the same training data we can construct the loss function
\begin{equation*}
    \hat\ell^{\text{win}}(\hat{y}) :=
    \frac{1}{N}\frac{1}{K}\sum_{i,k}\frac{1}{\Abs{J_{ik}}}\sum_{j \in J_{ik}}{
        \Norm{h(\xi_{ij}) - \hat{y}(t_{ij} - t_{ik}, \xi_{ik}, \sigma^{t_{ik}}u_{i})}_1
    },
\end{equation*}
where each $J_{ik} \subset [k, K]$, and $\Abs{J_{ik}}$ denotes the cardinality of the set.
In this case, the maximum length of the input sequences is given by
${L := 1 + \max_{i, k}\max_{j \in J_{ik}} \lfloor{\frac{t_{ij} - t_{ik}}{\Delta}}\rfloor}$,
which we can choose by appropriate selection of the index sets $J_{ik}$.
This allows for controlling the time complexity of the training epochs
and the smoothness of the loss function.
Of course, it is not necessarily the case that $\hat\ell^{\text{win}}$
is the empirical mean approximation of $\ell_T$, so care must be taken to avoid overfitting.

\section{Numerical experiments}\label{sec:experiments}
We perform two experiments with data from simulations of two conductance-based models:
a single neuron model,
and a model of the feedforward interconnection of two neurons with an electrical synapse
\footnote{The source code used to run the experiments is available at \url{https://github.com/mcpca/flow-learning}.}.

\subsection{Model of a single fast-spiking neuron}\label{ssec:single-neuron}

\begin{figure}[htbp]
    \centering
    \begin{subfigure}[b]{0.49\textwidth}
        \centering
        \includegraphics[width=0.99\textwidth]{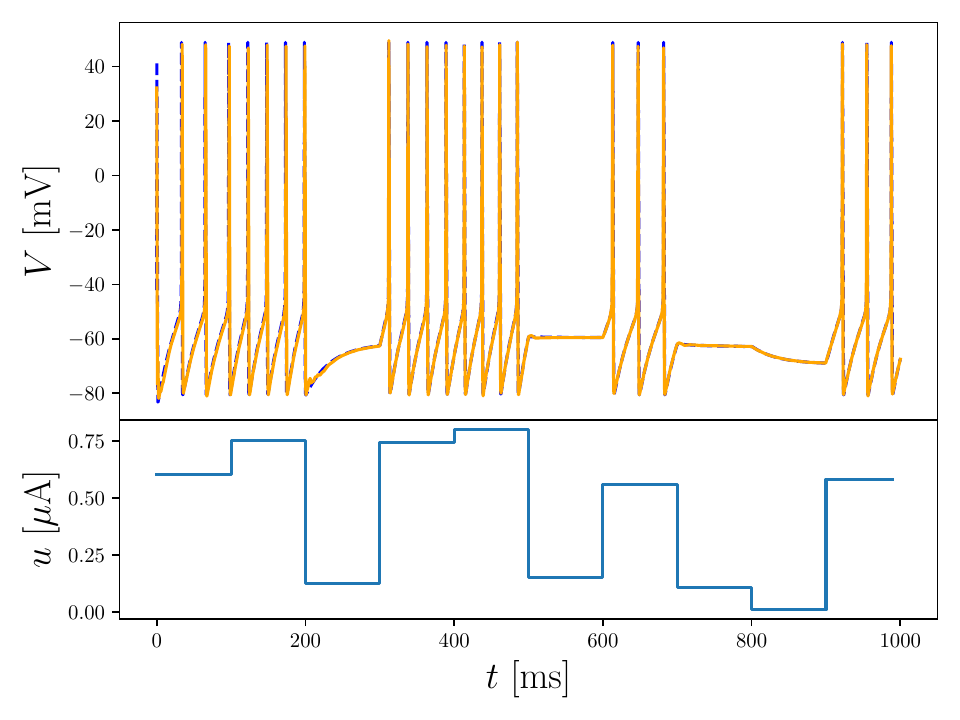}%
    \end{subfigure}%
    \hfill%
    \begin{subfigure}[b]{0.49\textwidth}%
        \centering
        \includegraphics[width=0.99\textwidth]{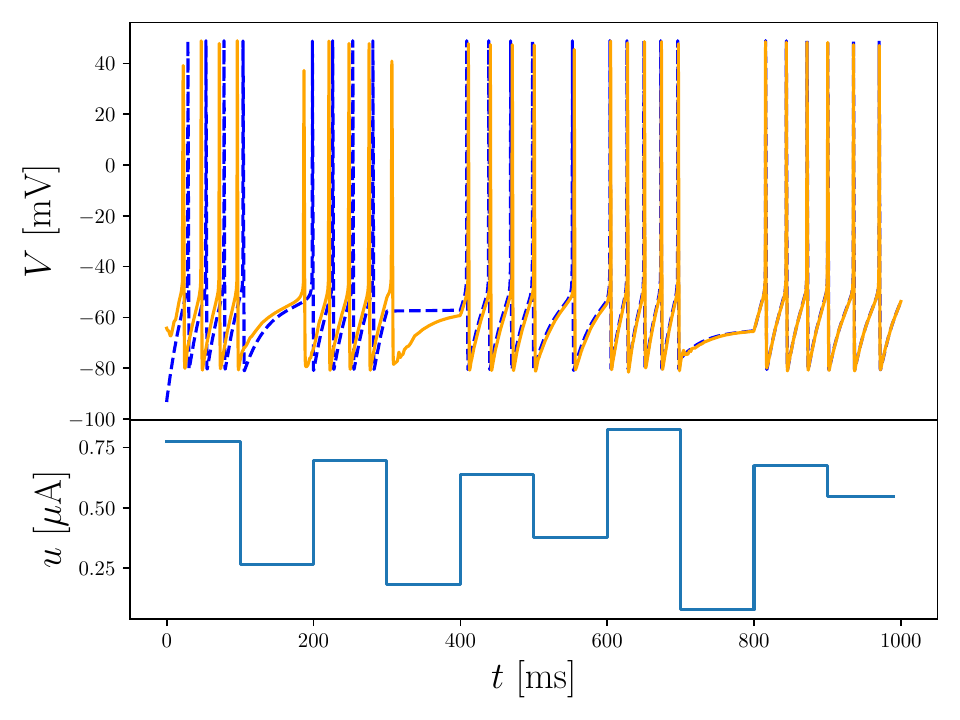}%
    \end{subfigure}
    \vspace{-1em}
    \caption{
        Predictions of output trajectories for a single fast-spiking neuron.
        The surrogate model prediction is shown in orange; the real output is plotted in blue (dashed).
        Note that the surrogate model closely reproduces the spiking behaviour of the neuron
        model, and is able to recover from errors in the prediction of the initial condition.
    }
    \label{fig:hhfs-predictions}
\end{figure}

\paragraph{Dynamics}
We consider a conductance-based model of a single neuron as in~\eqref{eq:cond-model} with 
a sodium current, 
$I_1(V, m_1, n_1) = g_{1} m_1^3 n_1 (V - V_{1})$,
and a potassium ionic current, 
$I_2(V, m_2, n_2) = g_{2} n_2^4 (V - V_{2})$,
i.e. $N_I = 2$.
The model can be described with four states, since ${\alpha_2 = 0}$ and so $m_2$ can be ignored.
This corresponds to a fast-spiking neuron,
and is identical to the model structure originally proposed in~\citet{HodgkinHuxley52}.
The full equations and parameter values may be found in~\citet{GiannariAstolfi22}.

\paragraph{Data collection}
We integrate $N = 800$ state trajectories of the model over $t \in [0, T]$, ${T = 500}$~ms,
using a backwards differentiation formula integrator,
and collect $K = 50 000$ samples from each trajectory,
with $t_{ik}$ sampled uniformly using Latin hypercube sampling.
The initial conditions are sampled uniformly with
${V(0) \sim \Unif{[-100, 100]}}$~mV
and ${m_k(0), n_k(0) \sim \Unif{[0, 1]}}$.
The control inputs have period ${\Delta = 10}$~ms and input values are sampled according to
\vspace{-.7em}
\begin{equation*}
\begin{aligned}
    \omega_{10k} &\overset{\text{i.i.d.}}{\sim} \Unif{[0, 1]}~\mu\text{A}, \ k \geq 0 \\
    \omega_{10k + j} &= \omega_{10k}, \ k \geq 0, \ 0 \leq j < 10,
\end{aligned}
\vspace{-.5em}
\end{equation*}
i.e. the input changes every 100~ms.

We reduce the dataset using the rejection sampling method described above, sampling according to the density
$p(t, x, u) \propto  \tilde{y}(t, x, u)$,
where $\tilde{y}$ is the normalisation of the output $y$ to $[0, 1]$.
We ensure that the local maxima of the signal (i.e. the spike peaks) and the initial state
are included in the final dataset.
The resulting sampled trajectories are split into training, validation, and testing sets
according to a 60/20/20\% random split.

\paragraph{Architecture and training}
We train 10~models with the architecture described in~\ref{ssec:architecture}. 
Each RNN is a long short-term memory (LSTM) network with 24~hidden states.
The encoder and decoder are feedforward networks with {$\tanh$}~activations
and three hidden layers.
The encoder network maps the initial condition to the initial hidden state of the LSTM.
The cell state of the LSTM is always zero-initialised.

We apply the windowing technique described in section~\ref{sec:methodology},
where $J_{ik}$ has at most 5~elements drawn uniformly (without repetition)
from $\Set{k, \dots, k + 20}$, so that the input sequences to the RNN have
maximum length $L = 20$.
The windowed empirical loss is minimised using the Adam algorithm with an initial learning rate 
of $1\times10^{-3}$.
The learning rate is reduced by a factor of~10 whenever the
empirical loss~\eqref{eq:emp-loss}~constructed with the validation data does not decrease for
5~consecutive epochs.
Training is stopped when the validation loss does not decrease for 15~consecutive epochs.

\begin{figure}[htbp]
    \centering
    \begin{subfigure}[b]{0.49\textwidth}
        \centering
        \includegraphics[width=0.99\textwidth]{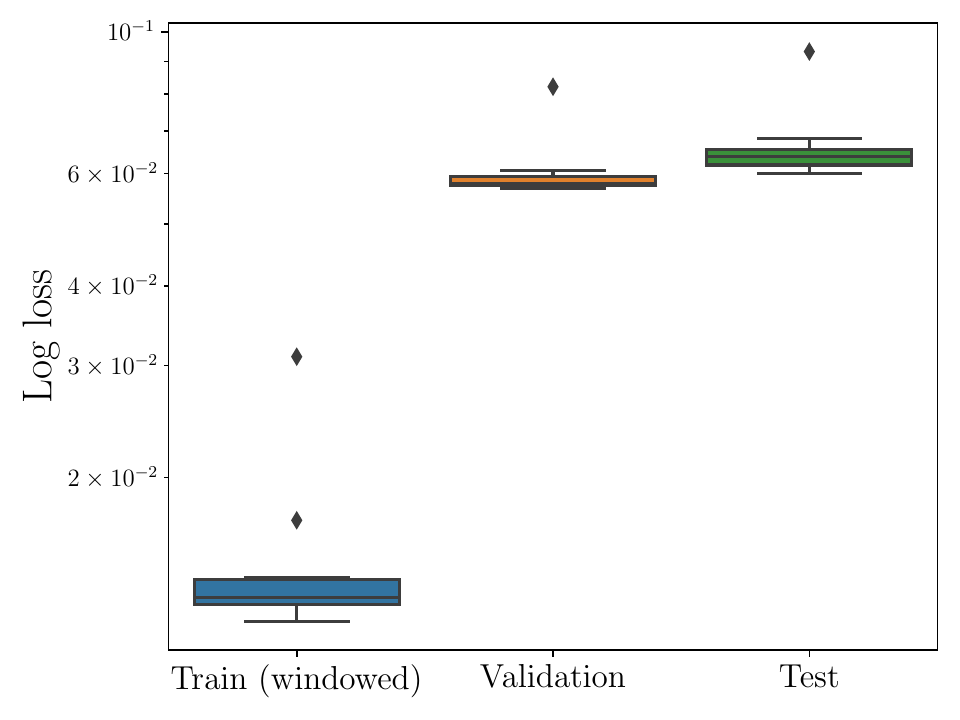}%
        \caption{}
        \label{fig:hhfs-losses}
    \end{subfigure}%
    \hfill%
    \begin{subfigure}[b]{0.49\textwidth}%
        \centering
        \includegraphics[width=0.99\textwidth]{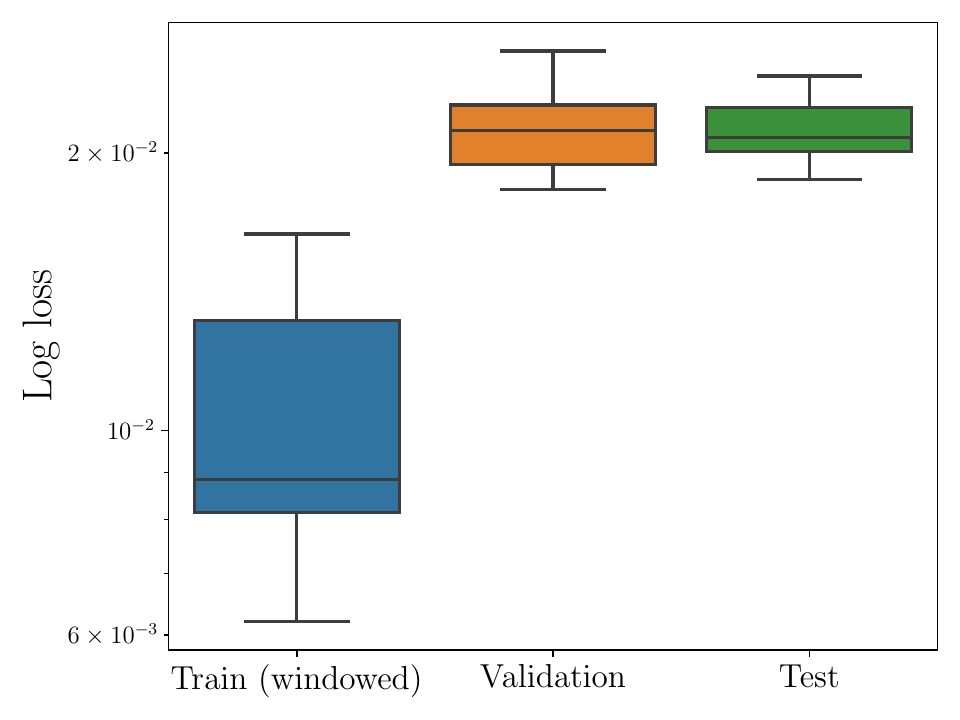}%
        \caption{}
        \label{fig:hhffe-losses}
    \end{subfigure}
    \vspace{-0.5em}
    \caption{
        Plot of the loss distributions for the 10 models trained with data from each system:
        (a) single fast-spiking neuron; (b) feedforward interconnection of two neurons.
    }
    \vspace{-2em}
\end{figure}

\paragraph{Results}
Figure~\ref{fig:hhfs-predictions} shows two trajectory predictions with 
unseen test inputs and initial conditions from the model with the
smallest validation loss among the 10 models.
We observe that the surrogate model is able to closely capture the timing
and height of the spikes.
In the right-hand side figure, we see that the model is not able to
predict the initial condition of the system correctly and consequently
misses the timing of the first few spikes, but nonetheless
correctly captures the spikes emitted after $t = 400$~ms.
It is interesting to note that although the system does not
have fading memory, i.e. the effect of initial conditions
does not necessarily disappear as $t \to \infty$,
the error in the initial conditions is not persistent
in the output of the surrogate model.
Figure~\ref{fig:hhfs-losses} shows the distributions 
of the losses for the 10 models, and we observe that the training procedure is
robust to the initialisation of the network parameters.

\subsection{Feedforward interconnection of two neuron models}\label{ssec:hhffe}

\paragraph{Dynamics}
We consider a model of the form~\eqref{eq:cond-model-multi} with $n = 2$.
Each of the neurons has $N_{I} = 3$ with $I_1, I_2$ as in the previous section 
and an additional potassium current given by
$I_3(V, m_3, n_3) = g_{3} n_3 (V - V_3)$,
so that each neuron has 5 states, and thus the interconnection results in a model with 10 states.
We take $\epsilon_{12} = 0.1$~S, $\epsilon_{21} = 0$~S, and $u_2 \equiv 0$,
corresponding to a feedforward interconnection of two regular spiking with adaptation type neurons,
as described in~\citet{GiannariAstolfi22}.

\paragraph{Data collection}
We follow the procedure described in the previous subsection, collecting 200~trajectories on
the time interval $[0, 1000]$~ms, with the same distributions for the initial conditions of the
membrane voltages and the gating variables, and the same distribution for the current input $u = u_1$.
The rejection sampling is performed with the density
$p(t, x, u) \propto \max_{i = 1, 2} \tilde{y}_i(t, x, u)$.

\paragraph{Architecture and training}
The details of the architecture and training procedure are as in the previous section,
the sole difference being that the LSTM network now has 32~hidden states.

\begin{figure}[htbp]
    \centering
    \begin{subfigure}[b]{0.49\textwidth}
        \centering
        \includegraphics[width=0.99\textwidth]{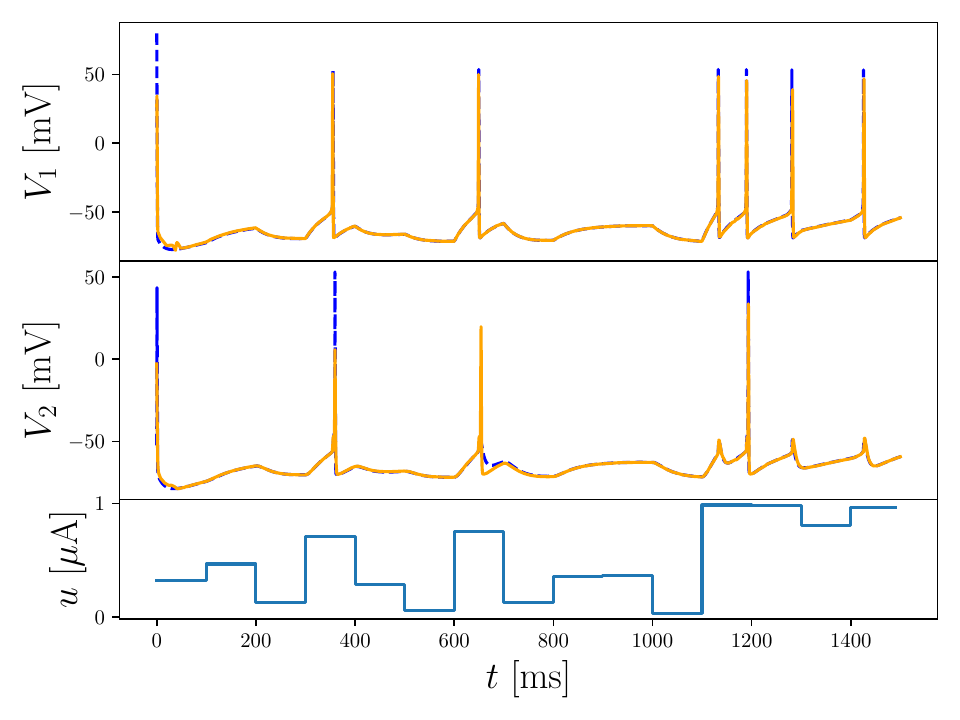}%
    \end{subfigure}%
    \hfill%
    \begin{subfigure}[b]{0.49\textwidth}%
        \centering
        \includegraphics[width=0.99\textwidth]{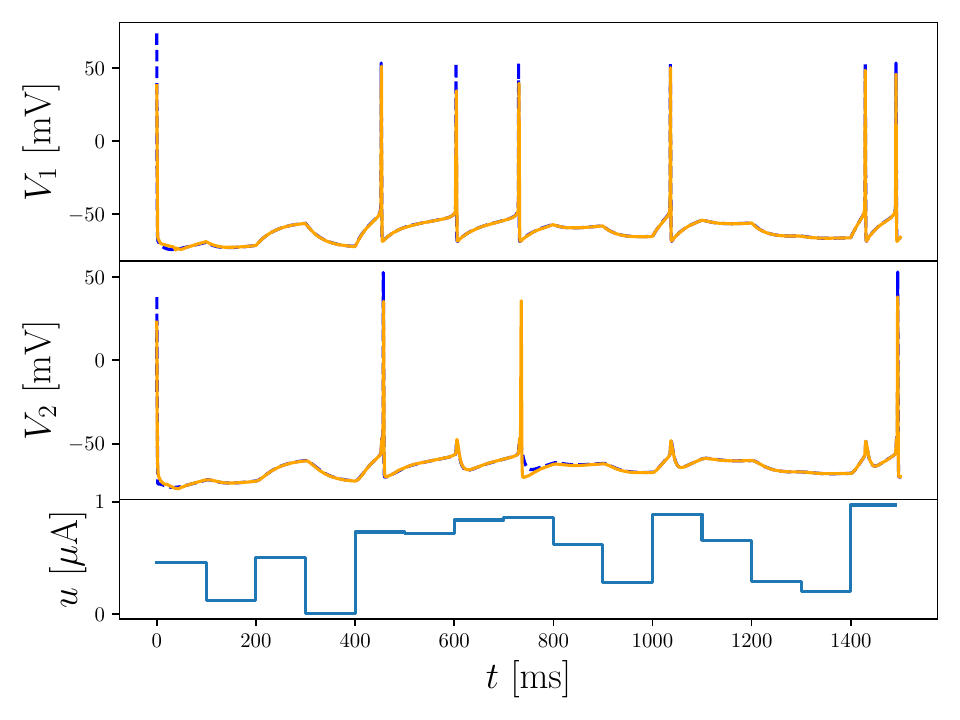}%
    \end{subfigure}
    \vspace{-1em}
    \caption{
        Predictions of output trajectories for the feedforward interconnection of two neurons.
        The surrogate model prediction is shown in orange; the real output is plotted in blue (dashed).
    }
    \label{fig:hhffe-predictions}
    \vspace{-2em}
\end{figure}

\paragraph{Results}
Figure~\ref{fig:hhffe-predictions} shows two trajectory predictions with 
unseen test inputs and initial conditions.
As in the previous subsection, we observe that the surrogate model faithfully
reproduces the behaviour of the spiking system.
Similarly, in Figure~\ref{fig:hhffe-losses} we verify the robustness
of the training procedure with respect to the training parameters.

\section{Conclusion}\label{sec:conclusion}

We proposed a framework for surrogate modelling of spiking systems based on
approximating the flow function of a class of state-space models exhibiting spiking behaviour.
The flow function approximation was performed using an RNN architecture which
allows for a direct continuous-time parameterisation of the output trajectories.
We discussed two issues which arise when training this architecture on data from a spiking
system, namely, the amount of data required to accurately represent the spike signals and
the complexity of the optimisation problem, and show how these can be addressed in the
context of our method.
Finally, we presented results from two numerical experiments which illustrate the feasibility
of using our framework for constructing surrogate models of spiking systems.

Directions for future research include an extended experimental study to validate the
applicability of the architecture in large networks of spiking systems,
and the possibility of further adapting the architecture to enforce 
system-theoretic properties characteristic of spiking systems.
It is also interesting to consider whether surrogate models of complex networks can be obtained
in a modular fashion.

\acks{
This work was supported by the Swedish Research Council Distinguished Professor
Grant~2017-01078 and a Knut and Alice Wallenberg Foundation Wallenberg Scholar Grant.
The computations were enabled by resources provided by the National Academic Infrastructure for Supercomputing in Sweden (NAISS) at C3SE partially funded by the Swedish Research Council through grant agreement no.~2022-06725.
}

\bibliography{bibliography}

\end{document}